# Representations in vision and language converge in a shared, multidimensional space of perceived similarities

**Katerina Marie Simkova (katerina.m.simkova@dartmouth.edu)**

Department of Psychological and Brain Sciences, Dartmouth College

Hanover, NH, USA

**Adrien Doerig (adrien.doerig@fu-berlin.de)**

Department of Education and Psychology, Freie Universität Berlin,

Berlin, Germany

**Clayton Hickey (c.m.hickey@bham.ac.uk)**

CHBH, School of Psychology, University of Birmingham

Birmingham, England, United Kingdom

**Ian Charest (ian.charest@umontreal.ca)**

cerebrUM, Département de Psychologie, Université de Montréal

Montréal, Québec, Canada

# Abstract

Humans can effortlessly describe what they see, yet establishing a shared representational format between vision and language remains a significant challenge. Emerging evidence suggests that human brain representations in both vision and language are well predicted by semantic feature spaces obtained from large language models (LLMs). This raises the possibility that sensory systems converge in their inherent ability to transform their inputs onto shared, embedding-like representational space. However, it remains unclear how such a space manifests in human behaviour. To investigate this, sixty-three participants performed behavioural similarity judgements separately on 100 natural scene images and 100 corresponding sentence captions from the Natural Scenes Dataset. We found that visual and linguistic similarity judgements not only converge at the behavioural level but also predict a remarkably similar network of fMRI brain responses evoked by viewing the natural scene images. Furthermore, computational models trained to map images onto LLM-embeddings outperformed both category-trained and AlexNet controls in explaining the behavioural similarity structure. These findings demonstrate that human visual and linguistic similarity judgements are grounded in a shared, modality-agnostic representational structure that mirrors how the visual system encodes experience. The convergence between sensory and artificial systems suggests a common capacity of how conceptual representations are formed—not as arbitrary products of first order, modality-specific input, but as structured representations that reflect the stable, relational properties of the external world.

# Introduction

Humans effortlessly translate between what they see and what they say, suggesting that our brains encode the meaning of a scene in a unified format. This intuition has led to the hypothesis of a shared high-level representational space for vision and language – a common code in which conceptual content is expressed regardless of whether it enters through the eyes or through words (Doerig et al. 2024). Establishing the existence of such shared representations is crucial for understanding how disparate streams of information coalesce into unified cognition. However, demonstrating a shared structure across modalities poses a significant challenge – especially when using rich, naturalistic inputs. Real-world scenes and narrative descriptions are complex and unconstrained, making it difficult to directly compare their internal representations.

Recent advances suggest that large language model (LLM) embeddings could constitute such a representational format. Emerging evidence in cognitive computational neuroscience suggests that LLM embeddings of natural scene captions are well aligned with brain responses elicited when individuals view corresponding natural scene images (Doerig et al., 2024; Lahner et al., 2024; Tang et al., 2023). For example, Doerig et al., 2024 found that training visual Deep Neural Networks (DNNs) to predict LLM embeddings leads to state-of-the-art models of visually evoked neural activities. In addition, aligning visual and linguistic representations when modelling natural scenes has been shown to provide even better predictors of visual responses in the brain (Conwell et al., 2022, Conwell et al., 2023, Wang et al., 2023), and to improve performance in multiple computational tasks (Lu et al. 2019; Tan and Bansal 2019; Pramanick et al. 2022; Radford et al. 2021). One explanation for the success of LLMs in predicting the brain's visual responses is that, through training on vast linguistic corpora, LLMs learn to capture rich statistical world-knowledge (Kaplan et al. 2020; Hernandez et al. 2021). Similarly, the brain needs to extract rich representations infused with world-knowledge from visual inputs to achieve a robust understanding of the visual world. For this reason, both systems converge onto a similar representational format that accounts for the complex information conveyed in natural scenes (Huh et al. 2024).

These recent developments in the cognitive computational neuroscience of vision are parallelled with breakthroughs in language research, where growing evidence shows that LLMs are good predictors of natural language representations in the brain, both for visual reading inputs and auditory speech inputs (Russo et al. 2022; Caucheteux, & King, 2022; Dong, & Toneva, 2023; Goldstein et al., 2021; Jain, & Huth, 2018; Toneva, & Wehbe, 2019; Wehbe et al., 2014). Some evidence suggests that the feature space of LLMs even aligns with predictive processes (Schrimpf et al., 2021; see Tuckute et al., 2024 for a systematic review), and high-level language comprehension of narratives in the human brain (Caucheteux et al., 2022). More recently, (Nikolaus et al. 2024) showed that encoding models trained on visual brain activity generalise to activity in natural language comprehension and vice versa. This converging evidence from language and vision motivates the possibility that various modalities all transform their low-level sensory inputs into a shared representational format that is captured by large language models. This representational format could satisfy the challenging requirements of accommodating varying, naturalistic sensory inputs, capturing information expressive enough to be relevant for behaviour, and compressing information.

Here, to test whether vision and language share a common representational format well captured by LLM embeddings, we collected similarity judgements for natural scene images and for corresponding captions describing the images. We used representational similarity analysis (RSA) to quantify the alignment between the similarity judgements in behaviour. To anticipate, we show that natural scene images and sentence descriptions of these images share a similar representational space in behaviour. Both sets of similarity judgments were used to model visual representations in the human brain and were compared to LLM embeddings of scene captions, as well as to activations from state-of-the-art recurrent convolutional neural network (RCNN) models trained either to predict category labels or to match LLM embeddings of scene captions (Doerig et al. 2024). This revealed a high-level network of brain areas that appear to underlie the shared representational space, and showed that RCNNs trained to predict LLM embeddings closely approximate this network's structure. By examining how these behavioural representations align with neural and computational models, we provide new evidence for a shared representational structure in the human mind.

# Results

Sixty-three participants performed a multiple arrangements task (MA; [Kriegeskorte & Mur 2012](#)) on stimuli obtained from the Special100 database of the Natural Scenes Dataset (NSD; [Allen et al., 2022](#)). These 100 natural scene images, each with a corresponding sentence caption describing the image, were selected to maximise semantic diversity (see [Allen et al., 2022](#) for sampling details). In arrangement of the natural scene images (henceforth "visual modality"), participants viewed the entire set of images on a computer screen in a white arena (Fig. 1) and used a computer mouse to arrange the images according to their similarity, positioning the images such that the distance between them reflects their similarity. In arrangement of sentence captions (henceforth "linguistic modality"), participants followed the same procedure, but each sentence initially appeared as an asterisk with the text revealed only when the asterisk was selected via mouse cursor. To control for the cognitive load between the visual and linguistic modalities, we also introduced a control condition where participants completed the MA task on "hidden images" (Supplementary Figure 1). Here, the same set of natural scene images was used, but each image was initially displayed as a white rectangle appearing only after it was selected via mouse cursor. The MA tasks adopted the 'lift-the-weakest' algorithm, which selected a subset of stimuli to be presented in subsequent trials to correct for placement errors ([Kriegeskorte & Mur 2012](#)). The trials terminated once 45 minutes elapsed, creating a final modality-specific RDM for each participant. To prevent task order effects from contaminating the representational geometry across modalities, the task order was counterbalanced across participants, with each attending three experimental sessions scheduled at least a week apart.

**Uncovering a shared similarity space between visual and linguistic stimuli**

To investigate the overall overlap between the similarity judgements in vision and language, we opted for the fixed-effects RSA approach in which we averaged the modality-specific RDMs across all participants and compared them using Spearman correlation. We found a strong relationship between the visual and linguistic RDMs (Figure 1B, fixed effects Spearman ρ = 0.781). A similar effect was observed when we accounted for inter-individual variability using the random-effects approach, where we computed pairwise correlations across all individuals in the visual and linguistic MAs (random effects Spearman ρ = 0.161, $p < 0.0001$). It is worth noting that we observed significant overlap between the modalities in initial experimental sessions where neither of the groups were familiar with the content of the other modality (Supplementary Figure 1). This demonstrates that even simple sentence descriptions convey enough information for the similarity relations to align with those of natural scene images.

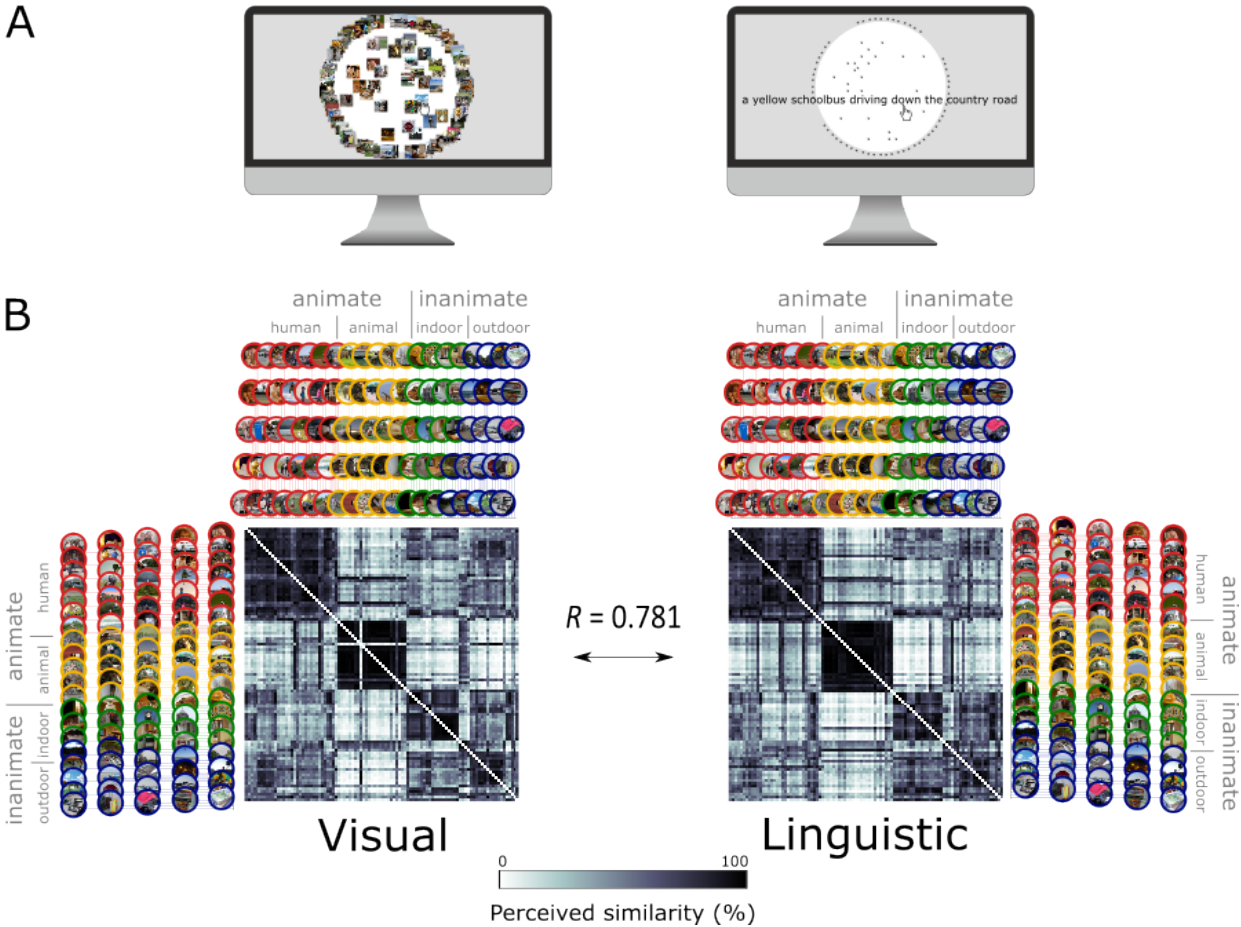

**Figure 1. Experimental design of the MA tasks.** (A) Participants completed the MA task either on 100 natural scene images (visual modality left) or 100 sentence captions

describing the images (linguistic modality right). To handle the large set of sentence captions, we adapted the MA method such that every caption was depicted as an asterisk and only the item currently being dragged was displayed. (B) The rank transformed subject-averaged modality-specific RDMs across all sessions were correlated to investigate the representational overlap between the visual and linguistic similarity judgements. We averaged modality-specific RDMs across participants before calculating Spearman correlation of these results, which generated a correlation of 0.781, suggesting strong representational overlap. The icons are colour-coded to differentiate between categories (red - human, yellow - animal, green - indoor, blue - outdoor).

## Visual and linguistic similarity spaces predict visually evoked brain activity in areas aligned with LLM embeddings of scene captions

The results above show that similarity judgements in vision and language lead to strikingly similar representational geometries in behaviour, but they leave unclear whether this reflects similar brain representations. To address this gap, we used the visual and linguistic similarity judgements to model visual responses in the human brain. We used the publicly available Natural Scenes Dataset ([Allen et al., 2022](#)), which consists of high-field (7T) functional MRI data from 8 participants who completed a visual recognition task involving the same set of 100 natural scene images used in our multiple arrangements task. We first extracted response patterns across the entire cortex using the searchlight procedure. We then trained a linear model using the behavioural MA RDMs to predict searchlight brain RDMs using cross-validated non-negative least squares regression (NNLS). The representational alignment was then quantified as the correlation between the behaviour-predicted RDMs and observed brain RDMs. Figure 2B shows that behaviour-predicted RDMs, whether based on visual (Figure 2B top panel) or linguistic similarity judgements (Figure 2B bottom panel), both exhibit a similar alignment with observed brain RDMs in regions located bilaterally along the mid- and high-level visual areas. The group-averaged Pearson correlations peaked at 0.363 for the visual modality and 0.353 for the linguistic modality. Results for each NSD participant are similar to group-level results (Supplementary Figure 2). These results show that similarity judgements of images and sentences align with much the same high-level visual brain regions.

Because participants completed similarity judgements on both images and sentences, it is possible that the alignment of similarity judgements across modalities stemmed from participants' familiarity with the stimuli. Specifically, participants may have intentionally arranged sentences in Session 3 much as they arranged images in the preceding session, potentially obscuring any differences between the modalities. To test whether the alignment between the behaviour-predicted and observed brain RDMs persists even before exposure to the other modality, we performed an additional analysis in which we predicted the brain RDMs using only behavioural RDMs from Session 1. The results revealed a significant representational alignment between the Session 1 behaviour-predicted RDMs and observed RDMs with correlation peaks in the occipitotemporal cortex (Supplementary Figure 3) similar to what was found in Figure 2. Participant-level analysis displayed similar predictions (Supplementary Figure 4). These results demonstrate that it is possible to reliably model visual responses using similarity judgements of sentence captions that were made by participants who had never seen the corresponding images before.

Alternatively, the representational alignment could be driven by the NNLS fitting procedure, which may bias the predicted RDMs toward an optimally weighted combination of behavioural participants, potentially inflating prediction accuracies. To rule out this possibility, we correlated the averaged modality-specific behavioural RDMs from Session 1 with observed RDMs at every searchlight. Supplementary Figures 5 and 6 reveal significant alignment in the occipitotemporal cortex for both visual and linguistic modalities, similar to results in Figure 2. This control analysis shows that the representational alignment is not merely an artifact of NNLS fitting. Notably, this holds despite using behavioural RDMs from Session 1, eliminating the possibility that familiarity played a role. Together, these findings support the idea that the representational geometry of visually evoked responses is shared with the psychological similarity structure of natural scene images and sentence captions.

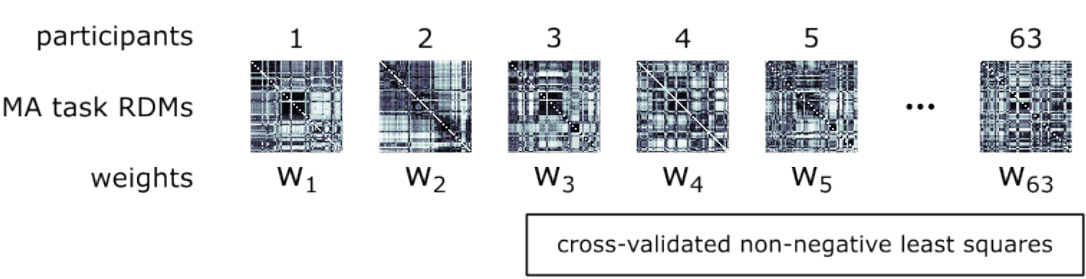

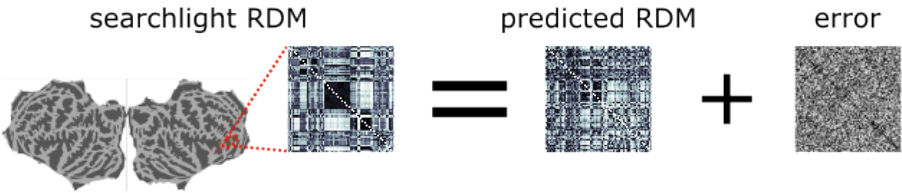

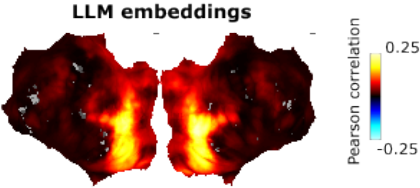

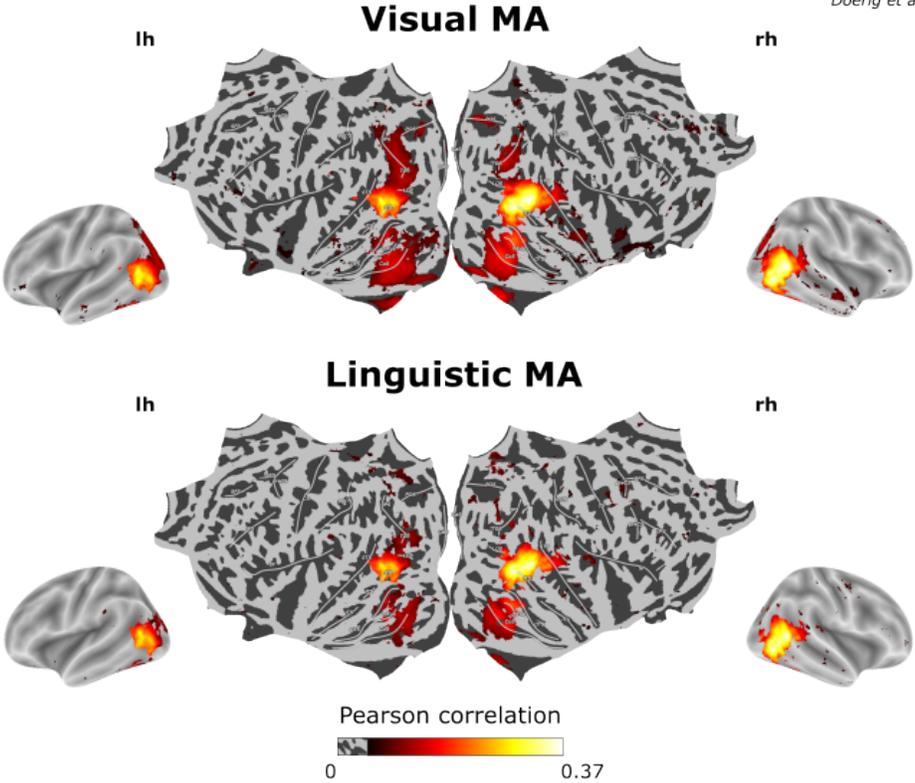

**Figure 2. Representational alignment of behaviour-predicted and observed brain RDMs in natural scene viewing.** (A) Cross-validated non-negative least squares regression was used to model the brain RDMs at every searchlight location using the behavioural RDMs derived from our MA tasks. (B) We averaged Pearson correlations across folds and determined significance across participants using a one-sided t-test (FDR corrected $p < 0.05$ level). The top panel displays alignment between observed brain RDMs and behaviour-predicted brain RDMs from the visual modality. The bottom panel displays alignment with behaviour-predicted brain RDMs from the linguistic modality.

## LLM-trained visual ANNs capture visual and linguistic similarity spaces

Which computational models best capture the structure of these visual and linguistic similarity spaces? To investigate this, we tested two types of visual Recurrent Convolutional Neural Networks (RCNNs) that take images as input: one was trained to predict scene category labels, while the other was trained to predict LLM embeddings of scene captions generated by MPNet (Doerig et al. 2024). This was motivated by a vast literature showing that object categories play a major role in ventral stream representations (Khaligh-Razavi and Kriegeskorte 2014; Kriegeskorte et al. 2008; Ungerleider and Haxby 1994; Ishai et al. 1999; Martin et al. 1996; Güçlü and van Gerven 2015; Doerig et al. 2023). However, recent evidence suggests that training ANNs to predict LLM embeddings of scene captions from visual inputs may be even more effective in modelling brain activities during natural scene viewing (Doerig et al. 2024). Aside from the training objective, the category- and LLM-trained RCNNs were identical in their architecture, training data, and random seeds. This ensured that any differences in their fit to human similarity judgements were driven solely by the training objectives.

We constructed RCNN RDMs by computing pairwise dissimilarities across activity patterns in each layer and time-point of the LLM-trained and category-trained models across 10 random seeds. We then created behaviour-predicted RDMs using the cross-validated NNLS procedure described earlier, predicting the RCNN RDMs from the behavioural MA RDMs of all subjects. To investigate which model, layer and timestep best captures the representational structure of human similarity judgements, we compared the behaviour-predicted RCNN RDMs to the observed RCNN RDMs either from the category- or LLM-trained RCNN at every layer and time step using Pearson correlation. We found that the behavioural RDMs predicted the representational structure of LLM-trained RCNNs significantly better than that of category-trained RCNN; this effect was observed for both visual and linguistic modality from layer 4 onward (Figure 3E, 3F). Behaviour-predicted RDMs also aligned significantly better with later time steps within each layer, regardless of the RCNN type, but the opposite trend was observed for layers 6 and lower (Figure 3A-F, see Supplementary Figure 7 for within-RCNN comparisons across layers and timesteps). In control analysis we found that the LLM-trained RCNN outperformed Alexnet (Supplementary Figure 8). It also outperformed MPNet embeddings (i.e., the LLM embeddings used to train our LLM-trained RCNN) in the visual modality, though not in the linguistic modality (Supplementary Figure 9). Together, these findings show that LLM-trained RCNNs are markedly good at predicting similarity judgements, suggesting that the multidimensional, embedding-like structure of these models best reflects how humans encode knowledge about the world.

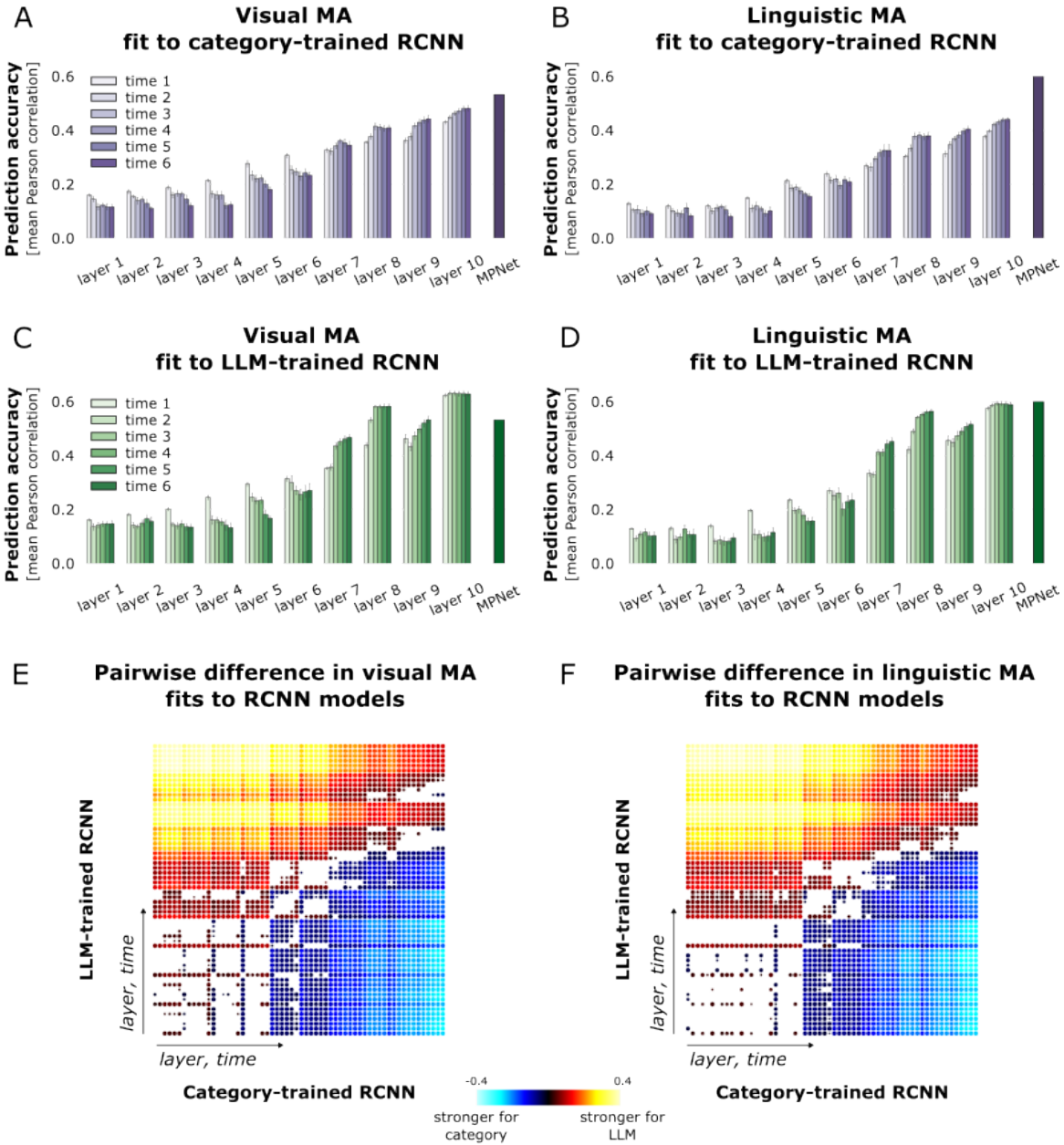

**Figure 3. Representational alignment between behaviour-predicted RCNN RDMs and observed RCNN RDMs.** Bars indicate Pearson correlation between behaviour-predicted RCNN RDMs and observed RCNN RDMs for each layer and time step, averaged across 10 models trained with different random seeds. Black bars represent Pearson correlation between the behaviour-predicted MPNet RDM and observed MPNet RDM, computed from the embeddings of scene captions. Error bars show the standard error of the mean, estimated across 10 cross-validation folds for each seed. Panels (A) and (B)

display prediction accuracies for category-trained RCNNs based on visual and linguistic MA, respectively. Panels (C) and (D) show prediction accuracies for LLM-trained RCNNs based on visual and linguistic MA, respectively. Pairwise difference in prediction accuracies between LLM- and category-trained RCNNs was calculated to identify a better fitting RCNN in the visual (E) and linguistic MA (F). Two-sided t-tests were performed on Pearson correlations across seeds for each layer and time step to assess significant differences between RCNN RDM fits to behaviour-predicted RDMs. Only statistically significant differences are shown. Point radius reflects p-value size ($p < 0.05$ FDR corrected), and colour gradients indicate the magnitude of the difference, calculated by subtracting category-trained from LLM-trained RCNN prediction accuracies. Layer and time complexity increase along both axes from the origin.

# Discussion

Humans can effortlessly describe what they see, and "see" what they read, but establishing a shared representational format between vision and language has been a challenging task. Here, we collected similarity judgements of natural scene images and associated sentence captions to investigate whether these modalities have a shared similarity structure. The results reveal that similarity judgements of sentences closely mirror those of images, and even reliably predict brain responses to natural scene viewing in an independent group of participants. Importantly, these brain predictions emerge even when the linguistic similarity judgements were made by individuals who had not yet seen the corresponding images. These findings suggest the existence of a shared representational format that arises from the mind's ability to recognise relational structure between concepts, independent of the first-order visual experience. Furthermore, a comparison with state-of-the-art computational models reveals that behavioural similarity judgements align most closely with the rich, semantic structure of LLM-based models, offering an insight into how concept representations are formed in the human mind.

The idea of second-order isomorphism has been considered as a core organisational principle of conceptual representations ([Shepard & Chipman, 1970](#)). This posits that the mind's internal representations preserve the relational structure of objects observed in the environment. While this account has been applied primarily to vision – examining how similarity judgements of visual objects relate to brain activity during object viewing ([Charest et al., 2014](#); [Cichy et al., 2019](#); [Contier et al., 2024](#); [Mur et al., 2013](#)) - its role in other modalities has remained underexplored. We show that a similar relational structure emerges for both linguistic and visual inputs. This finding is consistent with that of [Dima et al. (2024)](#) who showed overlap between similarity judgements of actions in videos and sentences, extending the early proposals of second-order isomorphism to language. It may be tempting to interpret the overlap in similarity judgements as evidence for amodal representations that emerge in supramodal brain regions and constitute modality-neutral conceptual information ([Fairhall & Caramazza, 2013](#)). However, our brain predictions do not immediately support this interpretation, showing that the relational structure closely resembles how visual experience is encoded in the visual system.

Importantly, we found that not only visual, but also linguistic similarity judgments predicted brain responses to natural scene images along the mid- and high-level visual regions. We found that this effect persisted even when participants had not seen the images before they judged the similarity of captions. We speculate that this alignment of linguistic similarity judgements and responses to natural scene images suggests that, while text and natural scene images have different visual features, they both ultimately project to the same abstract, high-level representational space. This interpretation is in line with recent work suggesting that high level visual cortex uses a highly abstract representational format well modelled by complex features such as object co-occurrences ([Bonner & Epstein, 2021](#)) or LLM embeddings ([Doerig et al., 2024](#)). It may be that the visual system translates sensory inputs into modality-agnostic representations that reflect stable, relational patterns observed in the real-world environment. Reactivation of this relational structure in language may be what enables participants to arrive at such an accurate representational geometry that resembles those obtained in the visual contexts. This also provides compelling context to studies from the congenitally blind who appear to acquire a wealth of knowledge about the visual appearance of objects through language alone ([Bedny et al., 2019](#); [Kim et al., 2019](#)) and represent this information in the visual cortex ([Bedny et al., 2011](#)). Altogether, this converging evidence suggests that what the visual system ultimately cares about is a stable, relational structure, and this promotes the alignment between language and vision.

Further analyses revealed that the representational structure encoded in the human mind also closely resembles the way several LLM-based computational models encode information. More specifically, we found that behavioural similarity judgements better predicted the feature space of LLM-trained RCNNs and LLM embeddings than that of category-trained RCNNs and Alexnet. This may be because category-trained RCNNs collapse the rich structure of the natural scene into fixed, orthogonal labels, which fails to capture the rich information upon which similarity judgements are based. Note that the category-trained RCNNs have the exact same architecture, training set and random seeds than the LLM-trained RCNNs, such that the advantage of the LLM-trained models cannot be explained by differences in architecture, initialization or training data (Doerig et al., 2024). Note also that the Imagenet-trained AlexNet is trained on more images than our LLM-trained RCNN, and that its late layers contain more units, such that the amount of training data and the dimensionality of layers does not favour the LLM-trained RCNN. The success of LLM-based models therefore appears to stem from their capacity to represent rich, semantic content through training on vast linguistic corpuses, capturing general statistical world knowledge about how concepts relate to one another (Doerig et al., 2024). Our findings suggest that the human mind follows similar principles to translate natural scenes into interpretable, behaviourally relevant dimensions as the LLM-based models. This supports the rapidly expanding interest in using such models to understand how humans encode and understand the world (Conwell et al., 2024; Lahner et al., 2024; Shoham et al., 2024).

We found that visual similarity judgements predicted visuo-linguistic LLM-trained RCNNs significantly better than purely linguistic LLM embeddings, whereas linguistic similarity judgements predicted both models equally well. This suggests that the shared similarity structure may retain traces of modality-specific information: the visual judgements likely capture perceptual detail that is not described in the captions, while the linguistic judgements draw more heavily on semantic content. This echoes recent findings that models jointly trained on visual and textual information outperform their uni-modal counterparts in many computational and predictive tasks (Andrews et al., 2009; Lazaridou et al., 2015; Wang et al., 2023). It suggests that vision and language each encode unique, yet complementary information that allows one to arrive at a richer representation of the visual world ([Collell & Moens, 2016](#); [Rubinstein et al., 2015](#)). The shared representational structure may therefore integrate features from both modalities. Future research should address the distinct representational dimensions projected from each modality to form a richer, more behaviourally relevant representation of the world.

In this study, we do not test how behavioural representational geometries align with neural activity during sentence reading. This leaves open the question of whether reading the captions would recruit the same neural populations that we identified in fMRI evoked by viewing the natural scene images. Further research will therefore require new MRI data collection. However, this matter has been partially addressed in recent work, in which participants either viewed natural scene images or read sentence captions during fMRI data collection ([Nikolaus et al., 2024](#)). This work has found that decoders trained on activity in high-level visual areas perform well both for decoding images and captions, indicating that patterns of brain activity elicited during sentence reading overlap with patterns elicited during scene viewing. Similar effects were observed during movie watching and story listening ([Popham et al., 2021](#); [Tang et al., 2023](#)). These findings suggest that future fMRI studies of sentence reading may uncover a similar predictive relationship between similarity judgments and activity patterns in mid- and high-level visual regions, as observed in the current study. Future research should address whether this relational similarity structure constitutes a universal principle of conceptual representations. If so, it may extend beyond vision to other domains such as sound or motion perception ([Barsalou, 2016](#)). Indeed, prior work has demonstrated that similarity among action-related concepts–whether conveyed through language or visual stimuli–can predict neural patterns in motor and premotor regions ([Hauk et al. 2004; Oosterhof et al. 2010](#)).

We draw several core observations from our results. First, these results support the idea that behavioural judgements of similarity in vision and language converge in the way they encode the relational structure of the physical world, independent of the first-order visual similarity. This relational structure closely mirrors how the visual system encodes sensory inputs in high-level visual areas. One exciting possibility is that this structure may reflect a general organizational principle for conceptual representations across modality-specific brain regions. Furthermore, we found that LLM-based models best accounted for the behavioral similarity structure, underscoring the richness and granularity of information they encode. Together, these findings lay the foundation for a shared relational space that captures behaviorally relevant structure across sensory modalities.

# Methods

## Participants

Sixty-six participants (23 ± 4 years of age) with normal or corrected-to-normal vision were recruited for the experiment. Three participants were excluded from analysis as they did not complete all three experimental sessions resulting in 63 participants left for analysis. Participants provided informed consent as part of a protocol approved by the University of Birmingham Research Ethics Committee and were awarded £8/h for participation.

## Stimuli

The stimulus set was obtained from the Special100 database of the Natural Scenes Dataset (NSD; [Allen et al., 2022](#)), and consisted of 100 natural scene images and 100 sentence captions describing the images. A greedy algorithm ([Allen et al., 2022](#)) selected the set of 100 images from a special set of 515 images such that they maximally span the high-level semantic space, which was defined by embeddings of the images' sentence captions. The sentence captions were collected from five human annotators as part of the Microsoft Common Objects in Context database ([Lin et al., 2014](#)). We identified the most frequent words and phrases across the five annotators and combined these to generate one representative sentence caption of every image.

## Procedure

Participants completed the visual, linguistic, and control MA tasks in three in-person sessions scheduled at least 7 days apart (average time between sessions 8.3 ± 3.3 days). The experiment was hosted by the web-based platform Meadows ([https://meadows-research.com/](https://meadows-research.com/)) and administered on a PC with a 25" monitor. The visual MA task (Fig. 1A, left) followed the standard procedure as introduced in [Kriegeskorte & Mur (2012)](#). Briefly, 100 natural scene images were displayed around a white circular arena.

Participants were then instructed to arrange the images inside the white arena according to their similarity using mouse drag-and-drop operations such that images perceived as more similar are placed closer together while images perceived as less similar are placed further apart. The linguistic MA task (Figure 1A, right) was conceptually similar, but involved the arrangement of sentence captions which allowed us to recover the representational geometry of the images' linguistic referents. To reduce the clutter introduced by the display of numerous lengthy sentences around the MA arena, we adapted the MA method such that every sentence was initially displayed as an asterisk. The sentence appeared only when the participant hovered the mouse cursor over the asterisk and sustained as the participant moved the location of this item. Other than that, the procedure of the linguistic MA was identical to the visual MA.

While the high-level representation is instantly accessible through direct perception in the visual modality, the representation in the linguistic modality is covert and relies on mental reconstruction of the sentence content. This introduced the possibility that the linguistic MA is cognitively more demanding than the visual MA task (i.e., reconstructing the mental image and recruiting working memory to preserve its representation). To control for the cognitive load of the overt visual and covert linguistic modality, we introduced a control condition in which the natural scene images were initially displayed as white rectangles (Supplementary Figure 1). As was the case for captions in the linguistic MA task, in the control task the images appeared only when participants hovered the mouse cursor over the white rectangle and as they moved the item in the arena. This method allowed us to employ the covert characteristic of the linguistic modality where participants are required to mentally reconstruct and maintain the representation. Other than that, the procedure was the same as in the remaining two tasks.

**Estimating the representational dissimilarity matrices**

The MA method used the weighted-averaging approach described in Kriegeskorte & Mur (2012) to obtain the representational dissimilarities. Following the first MA trial in which the full set of 100 stimuli was displayed, an algorithm computed Euclidean distances between every pair of stimuli ((100 items$^2$ - 100)/2 = 4950 pairs) inside the MA arena and assembled them in an initial RDM symmetric about a diagonal of zeros. Subsets of stimuli that showed the weakest weighted evidence within the RDM were then presented in subsequent MA trials. With every new trial, the representational dissimilarities in the initial RDM were iteratively adjusted by the weighted-averaging approach reducing potential placement errors. Because perceived dissimilarity between scenes is context dependent, this procedure also allowed us to account for the multidimensional nature of similarity relations. The MA trials terminated either when enough evidence was obtained for every pair of stimuli or once 45 minutes elapsed. This yielded three modality-specific RDMs per participant.

**Image-computable RCNN trained on category- and LLM embeddings of sentence descriptions**

The set of 100 natural scene images was passed through two recurrent convolutional neural networks (RCNNs) whose objective was to produce either a sentence caption or a category label of the image. The architecture of the network was introduced in detail in [Doerig et al., (2024)](). Briefly, the network was derived from vNet and was adapted to include both lateral and top-down recurrent connections. This architecture allowed us to investigate how behavioural RDMs align with recurrent activity found to correspond to representational dynamics of the visual system ([Kietzmann et al., 2019]()). As described in [Doerig et al., (2024)](), both RCNNs shared identical architecture and were trained with 10 random seeds on the MS COCO dataset with the NSD stimuli excluded. The only difference between the sentence- and category-trained RCNN is in the objective of the output layer. The sentence-trained RCNN minimised the cosine distance between the predicted and the target sentence embedding obtained from MPNet. The category-trained RCNN, on the other hand, used multi-hot encoding to predict the category label. After evaluating both RCNNs on the set of 100 natural scene images, we extracted sentence embeddings and category vectors for the last layer and time step from each of 10 random seeds. We calculated correlation distance between the output vectors from each RCNN and assembled the dissimilarity estimates in model RDMs.

**Natural Scenes Dataset**

A detailed description of the NSD (http://naturalscenesdataset.org) is provided elsewhere (Allen et al., 2022). The NSD dataset contains measurements of fMRI responses from 8 participants who each viewed 9,000–10,000 distinct color natural scenes (22,000–30,000 trials) over the course of 30–40 scan sessions. Scanning was conducted at 7T using whole-brain gradient-echo EPI at 1.8-mm resolution and 1.6-s repetition time. Images taken from the MS COCO database (Lin et al., 2014), square cropped, and presented at a size of 8.4° x 8.4°. The special set of 100 images used in our multiple arrangements task were shared across subjects. Images were presented for 3 s with 1-s gaps in between images. Subjects fixated centrally and performed a long-term continuous recognition task on the images. The fMRI data were pre-processed by performing one temporal interpolation (to correct for slice time differences) and one spatial interpolation (to correct for head motion). A general linear model was then used to estimate single-trial beta weights. Cortical surface reconstructions were generated using FreeSurfer, and both volume- and surface-based versions of the beta weights were created. In this paper, we used the 1.8-mm volume preparation of the NSD data and version 3 of the NSD single-trial betas (betas_fithrf_GLMdenoise_RR).

**Construction of searchlight brain RDMs**

Searchlight brain RDMs were constructed using NSD fMRI that were collected from 8 participants who viewed the set of 100 natural scene images in a recognition task (Allen et al., 2022). Activity patterns were extracted in a sphere with a radius of 6 voxels from each participants' single-trial betas. We then calculated pairwise correlation distances between the activity patterns and assembled them in a searchlight brain RDM.

**Cross-validated non-negative least squares procedure**

We created a behaviour-predicted RDM to investigate how well the behavioural geometry in vision and language aligns with patterns of visually evoked activity, and the representational structure of category- and sentence-computable RCNNs. Behavioural RDMs across all sessions were used unless stated otherwise. The NSD stimulus set was first sampled into train and test sets at 70:30 ratio using 10-fold cross-validation. For each fold, pairwise dissimilarities in behavioural and observed RDMs (either brain RDM or RCNN RDM) were split into the train and test sets. The train subset of behavioural RDMs was then fitted on the train subset of the observed RDM using non-negative least squares (NNLS) regression. A major advantage of this approach is that it allows us to assign flexible weights each representing a participant's relative fit to the train subset of the observed RDM. The NNLS weights were then used to create a behaviour-predicted RDM that is the best possible representation of the observed representational geometry. This was done using the dot product between the NNLs weights and the test subset of behavioural RDMs. We then evaluated the fit between the behaviour-predicted RDM and observed RDM using Pearson correlation.

      For searchlight brain RDMs, this procedure was repeated for each searchlight RDM and each NSD participant separately. Pearson correlations were averaged across folds for each NSD participant and projected from the original 1.8-mm volume to fsaverage using nsdcode (https://github.com/cvnlab/nsdcode/tree/master). For group statistics, Pearson correlations were averaged across NSD participants and activation maps were plotted using the pycortex (Gao et al., 2015) and nilearn (https://github.com/nilearn/nilearn) libraries. The significance of correlations was tested using one-sided t-test across participants and corrected for multiple comparisons at FDR $p < 0.05$. For RCNN comparisons, Pearson correlations were averaged across 10 seeds for each layer and time step and tested for significance using one-sided t-test corrected for multiple comparisons using FDR $p < 0.05$. To test the difference in correlations of category- and sentence-computable models with behaviour-predicted RDMs, Pearson

correlations were subtracted and tested for significance using one-sided t-test corrected at FDR p < 0.05.

**Acknowledgements**

The authors acknowledge support by the ERC stg grant 759432 START (Charest), a Courtois Chair in computational neuroscience (Charest), and an NSERC Discovery grant (Charest). Collection of the NSD dataset was supported by NSF IIS-1822683 (Kay) and NSF IIS-1822929 (Naselaris).

**Code and data availability**

The Natural Scenes Dataset is available at http://naturalscenesdataset.org. Behavioural data and code for the analyses reported here will be available upon publication of the manuscript.

Charest, I., Kievit, R. A., Schmitz, T. W., Deca, D., & Kriegeskorte, N. (2014). Unique semantic space in the brain of each beholder predicts perceived similarity. Proceedings of the National Academy of Sciences of the United States of America, 111(40), 14565–14570.

Cichy, R. M., Kriegeskorte, N., Jozwik, K. M., van den Bosch, J. J. F., & Charest, I. (2019). The spatiotemporal neural dynamics underlying perceived similarity for real-world objects. NeuroImage, 194, 12–24.

Collell, G., & Moens, M. F. (2016). Is an Image Worth More than a Thousand Words? On the Fine-Grain Semantic Differences between Visual and Linguistic Representations. Proceedings of COLING 2016, the 26th International Conference on Computational Linguistics: Technical Papers, 2807–2817.

Contier, O., Baker, C. I., & Hebart, M. N. (2024). Distributed representations of behaviour-derived object dimensions in the human visual system. Nature Human Behaviour, 1–15.

Conwell, C., Prince, J. S., Hamblin, C. J., & Alvarez, G. A. (2023). Controlled assessment of CLIP-style language-aligned vision models in prediction of brain & behavioral data. https://openreview.net/pdf?id=T90SJkeDKm

Conwell, C., Prince, J. S., Kay, K. N., Alvarez, G. A., & Konkle, T. (2022). What can 1.8 billion regressions tell us about the pressures shaping high-level visual representation in brains and machines? BioRxiv. https://www.biorxiv.org/content/10.1101/2022.03.28.485868.abstract

Conwell, C., Prince, J. S., Kay, K. N., Alvarez, G. A., & Konkle, T. (2024). A large-scale examination of inductive biases shaping high-level visual representation in brains and machines. Nature Communications, 15(1), 9383.

Neuroscience: The Official Journal of the Society for Neuroscience, 35(27), 10005–10014.

Hauk, O., Johnsrude, I., & Pulvermüller, F. (2004). Somatotopic representation of action words in human motor and premotor cortex. Neuron, 41(2), 301–307.

Hernandez, D., Kaplan, J., Henighan, T., & McCandlish, S. (2021). Scaling Laws for Transfer. In arXiv [cs.LG]. arXiv. http://arxiv.org/abs/2102.01293

Huh, M., Cheung, B., Wang, T., & Isola, P. (2024). The platonic representation hypothesis. In arXiv [cs.LG]. arXiv. http://arxiv.org/abs/2405.07987

Ishai, A., Ungerleider, L. G., Martin, A., Schouten, J. L., & Haxby, J. V. (1999). Distributed representation of objects in the human ventral visual pathway. Proceedings of the National Academy of Sciences of the United States of America, 96(16), 9379–9384.

Jain, S., & Huth, A. G. (2018). Incorporating context into language encoding models for fMRI. bioRxiv. https://doi.org/10.1101/327601

Kaplan, J., McCandlish, S., Henighan, T., Brown, T. B., Chess, B., Child, R., Gray, S., Radford, A., Wu, J., & Amodei, D. (2020). Scaling Laws for Neural Language Models. In arXiv [cs.LG]. arXiv. http://arxiv.org/abs/2001.08361

Khaligh-Razavi, S.-M., & Kriegeskorte, N. (2014). Deep supervised, but not unsupervised, models may explain IT cortical representation. PLoS Computational Biology, 10(11), e1003915.

Kietzmann, T. C., Spoerer, C. J., Sörensen, L. K. A., Cichy, R. M., Hauk, O., & Kriegeskorte, N. (2019). Recurrence is required to capture the representational dynamics of the human visual system. Proceedings of the National Academy of Sciences of the United States of America, 116(43), 21854–21863.
30

# Supplementary Figures

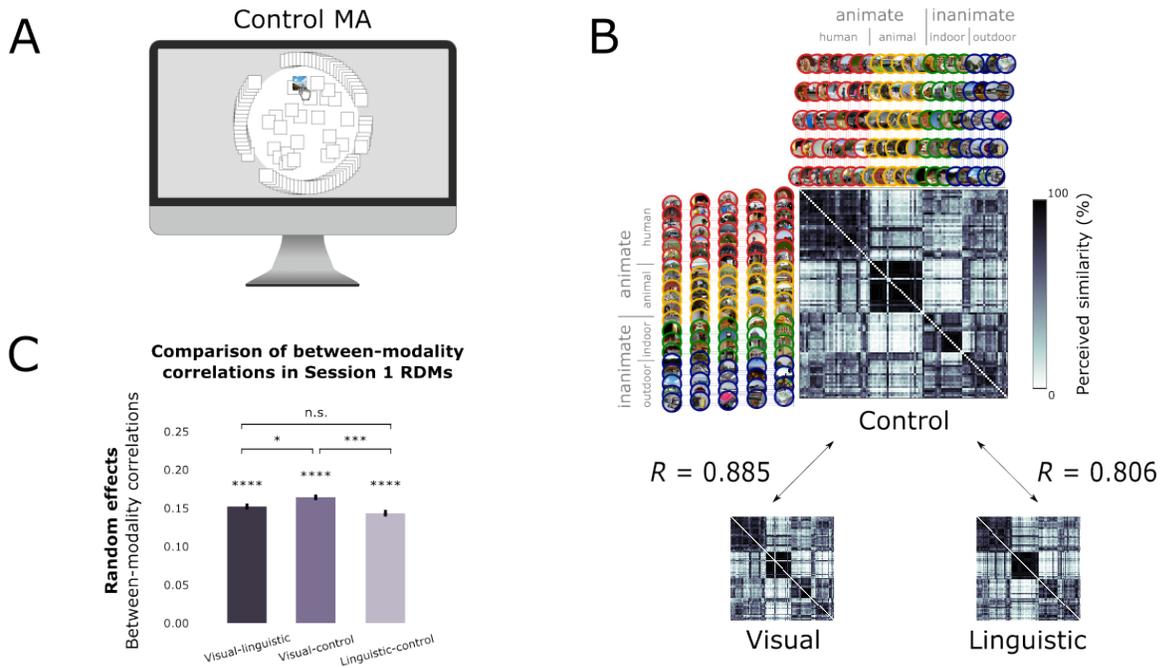

**Supplementary Figure 1. Experimental design of the control MA task and modality comparisons.** (A) In the control condition, participants were presented with the same set of 100 natural scene images, only these were displayed as white rectangles. The images appeared one at a time as the participant dragged or hovered over them with the mouse cursor. The same procedure was followed as in the other MA tasks. (B) RDMs across all sessions from the control condition were averaged and rank-transformed, and compared to the remaining two modalities. (C) The bars display average between-modality correlations across all pairwise RDM combinations (i.e., random effects analysis) for RDMs from Session 1 only. This enabled us to investigate the representational overlap before participants were familiar with the content of the other modalities than they started with. Independent samples t-test revealed that the control MA is significantly better aligned with the visual MA compared to the linguistic MA. No difference was found, whether the linguistic MA was correlated with the visual or control MA. The stars denote p-values with: *$p < .05$, ** $p < .01$, *** $p < .001$, **** $p < .0001$.

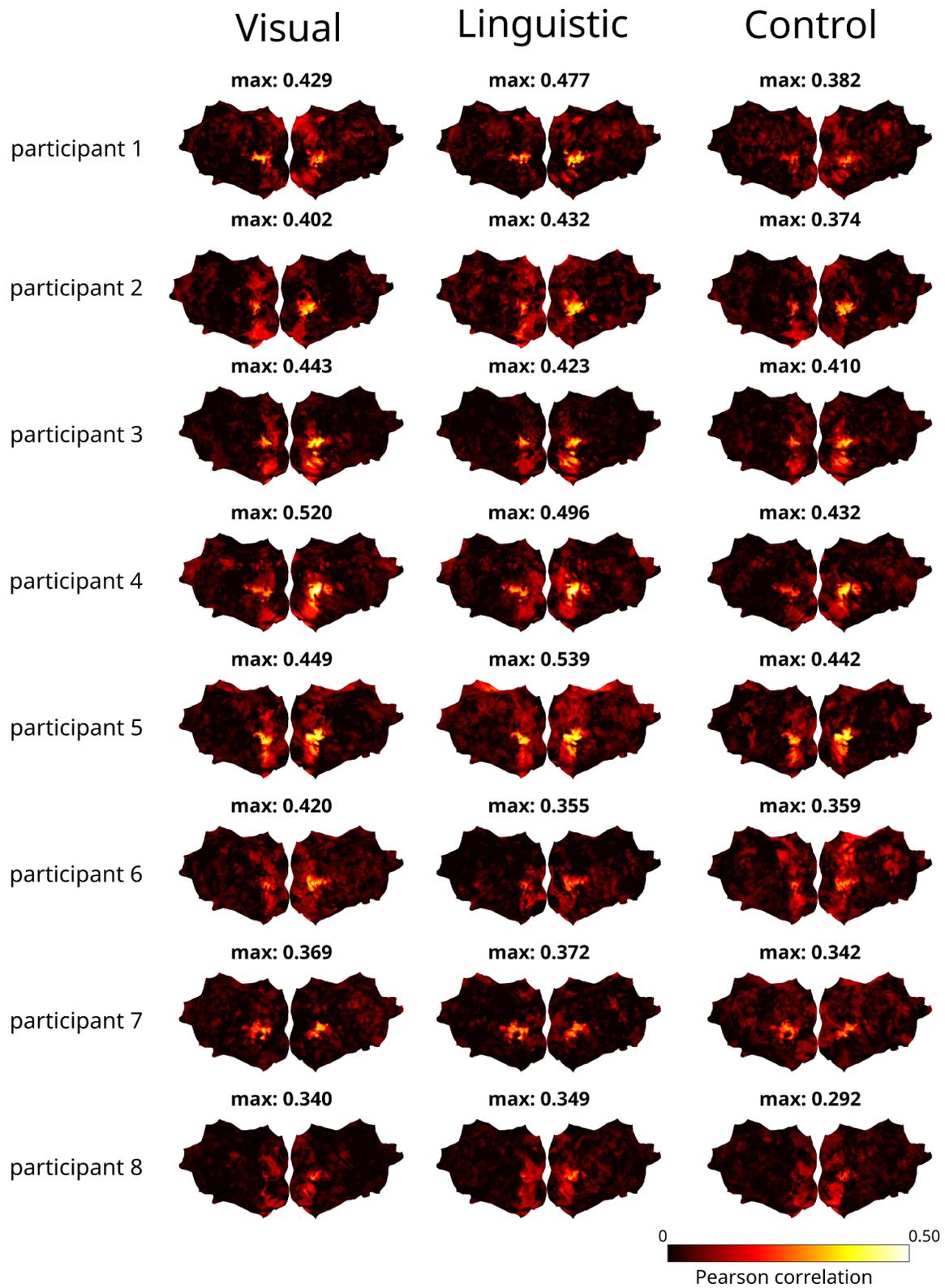

**Supplementary Figure 2. Representational alignment between the behaviour-predicted and observed brain RDMs across all sessions at the participant level.** For

each NSD participant, Pearson correlations between the behaviour-predicted and observed brain RDMs were averaged across 10 cross-validation folds. Representational alignment is reported for all three task modalities. Similar to group-level comparisons, the surface maps reveal strong and relatively stable correlations across all NSD participants and tasks bilaterally along the occipitotemporal cortex.

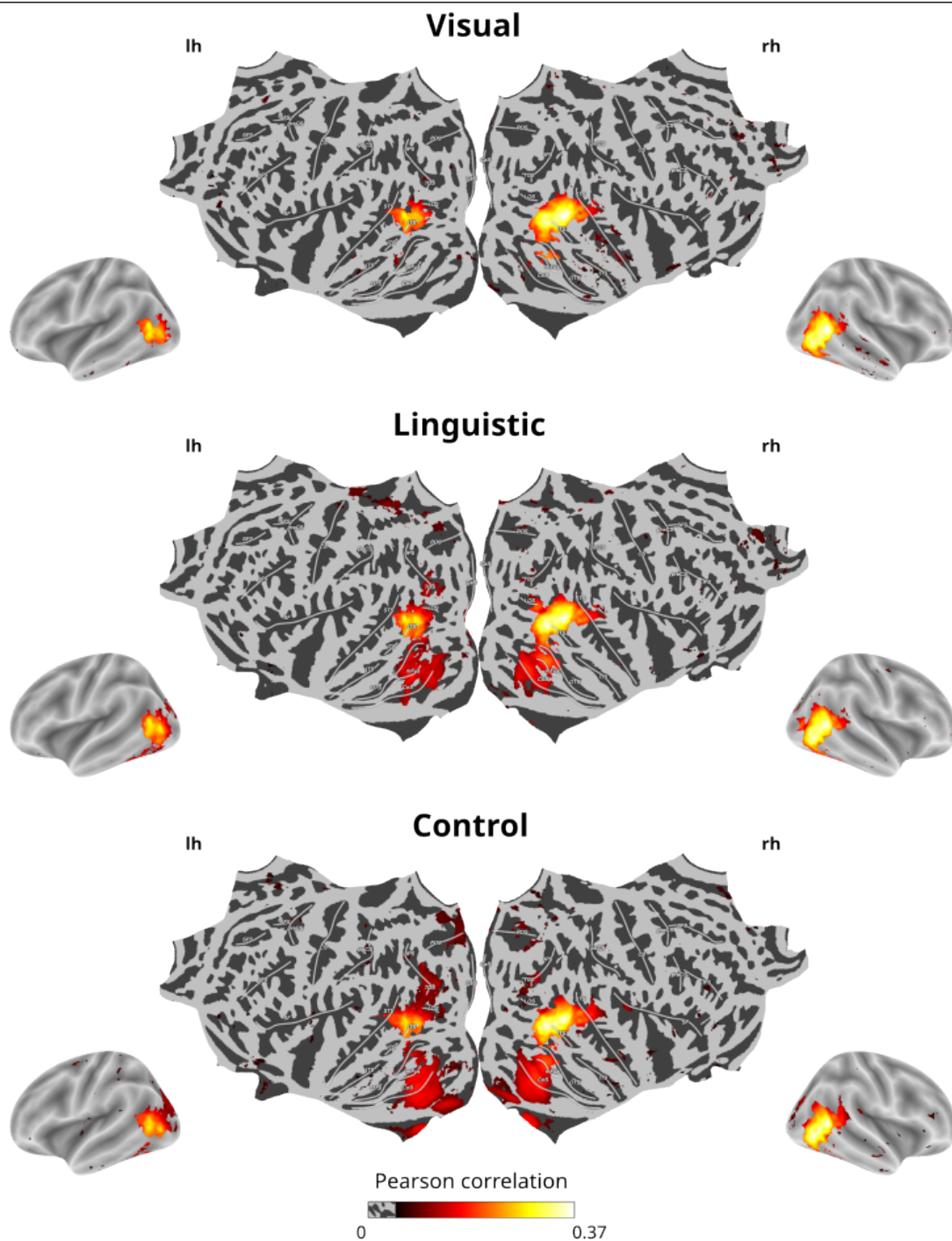

**Supplementary Figure 3. Representational alignment between the behaviour-predicted and observed brain RDMs in Session 1.** To demonstrate that representational alignment remains even before behavioural participants were familiar

with stimuli from the other MA task modality, only behavioural data from Session 1 were used to create the behaviour-predicted RDMs. The figure shows group-averaged Pearson correlations between the behaviour-predicted and observed brain RDMs ($p < 0.05$, FDR-corrected, one-sided test). The surface maps reveal a significant representational overlap between the RDMs spanning the occipitotemporal cortex similar to what was shown in Figure 2. The Pearson correlations peak at 0.347, 0.338, and 0.346 for visual, linguistic, and covert visual modality respectively. This result is reassuring because it demonstrates that even though behavioural participants had never seen the visual format of sentence captions in the linguistic MA before, they still reconstruct a representational geometry that is predictive of brain responses to viewing natural scene images.

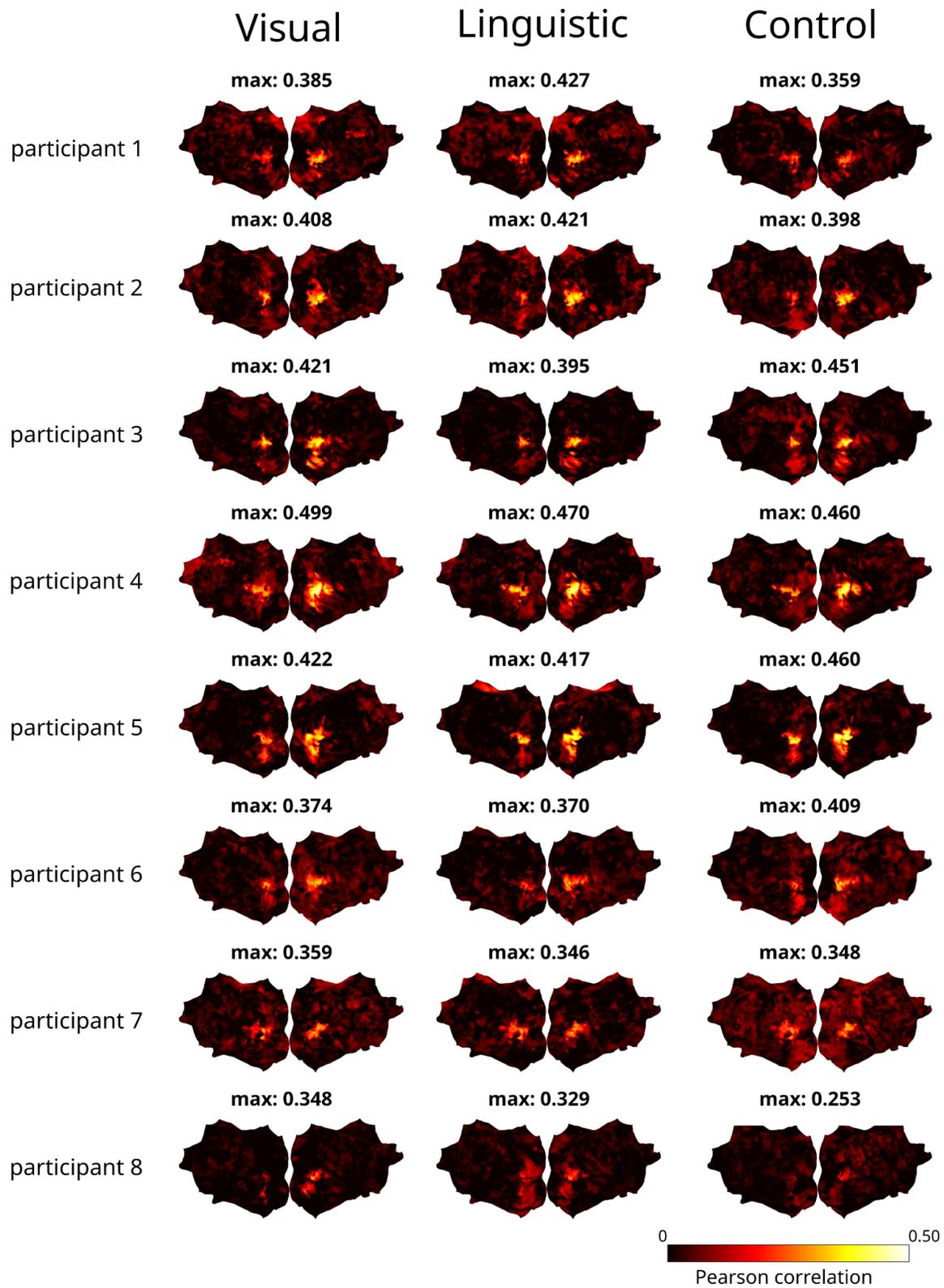

**Supplementary Figure 4. Representational alignment between the behaviour-predicted and observed brain RDMs in Session 1 at the participant level.** The figure

shows Pearson correlations between the behaviour-predicted and observed brain RDM averaged across folds for each NSD participant. Representational alignment is obvious across all three modalities despite only behavioural RDMs from Session 1 were considered here.

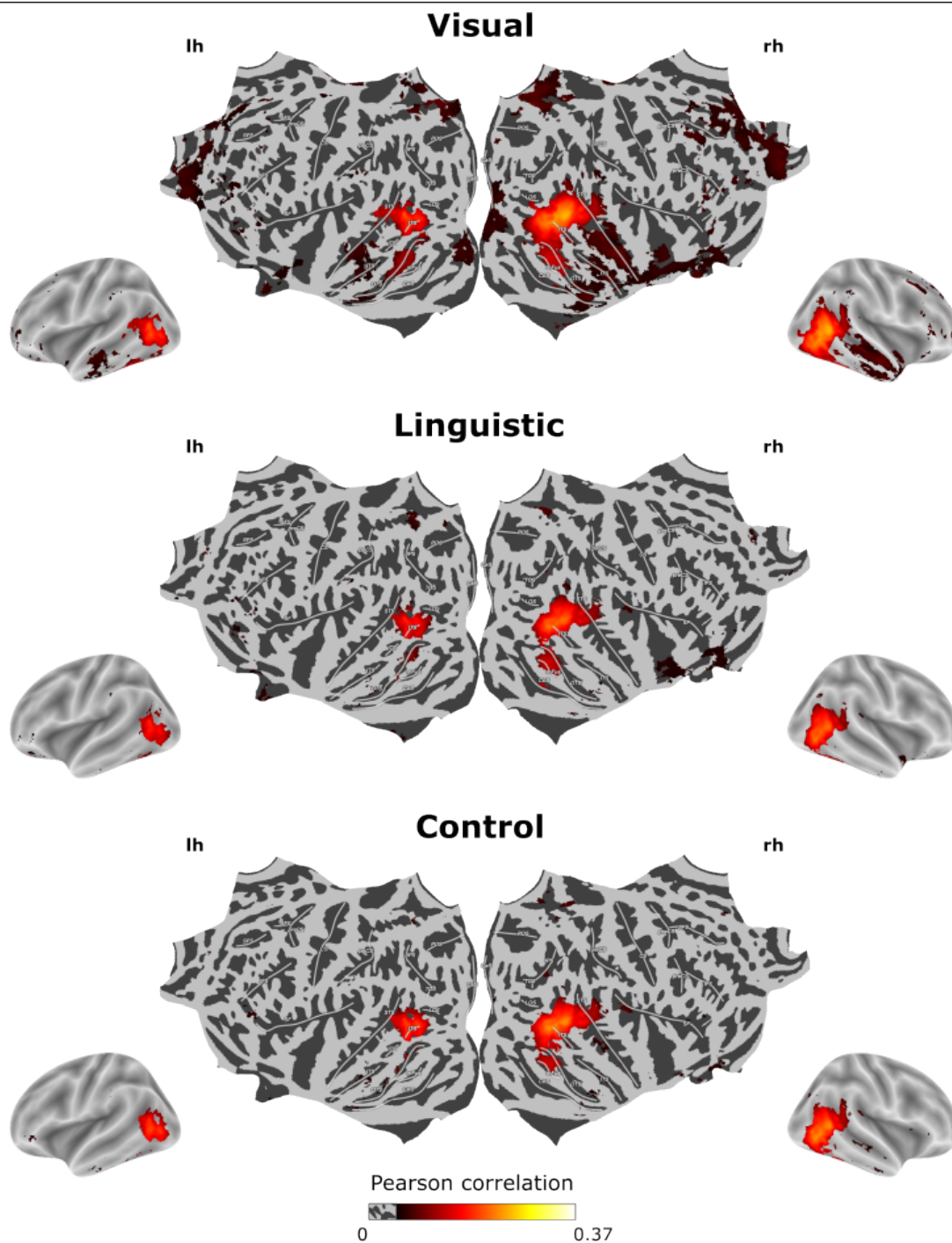

**Supplementary Figure 5. Representational alignment of behavioural similarity judgements in Session 1 and visually evoked brain responses.** To demonstrate that the representational alignment is reproducible even without creating a best fitting

behaviour-predicted RDM, we correlated the averaged modality-specific behavioural RDMs from Session 1 with brain RDMs at every searchlight for each NSD participant. The surface maps show averaged Pearson correlations across the NSD participants ($p < 0.05$, FDR-corrected, one-sided test). The correlations peaked at 0.239, 0.204, and 0.220 for visual, linguistic, and control modality respectively. The spatial distributions of the correlations on the cortex are similar to the ones reported in Figure 2B. This finding shows that the similar spatial distribution on the cortex observed with the linguistic and visual behavioural RDMs cannot simply be explained by the re-weighting procedure. This is even more significant considering that we used behavioural RDMs from Session 1 where no cross-modal projection played a role.

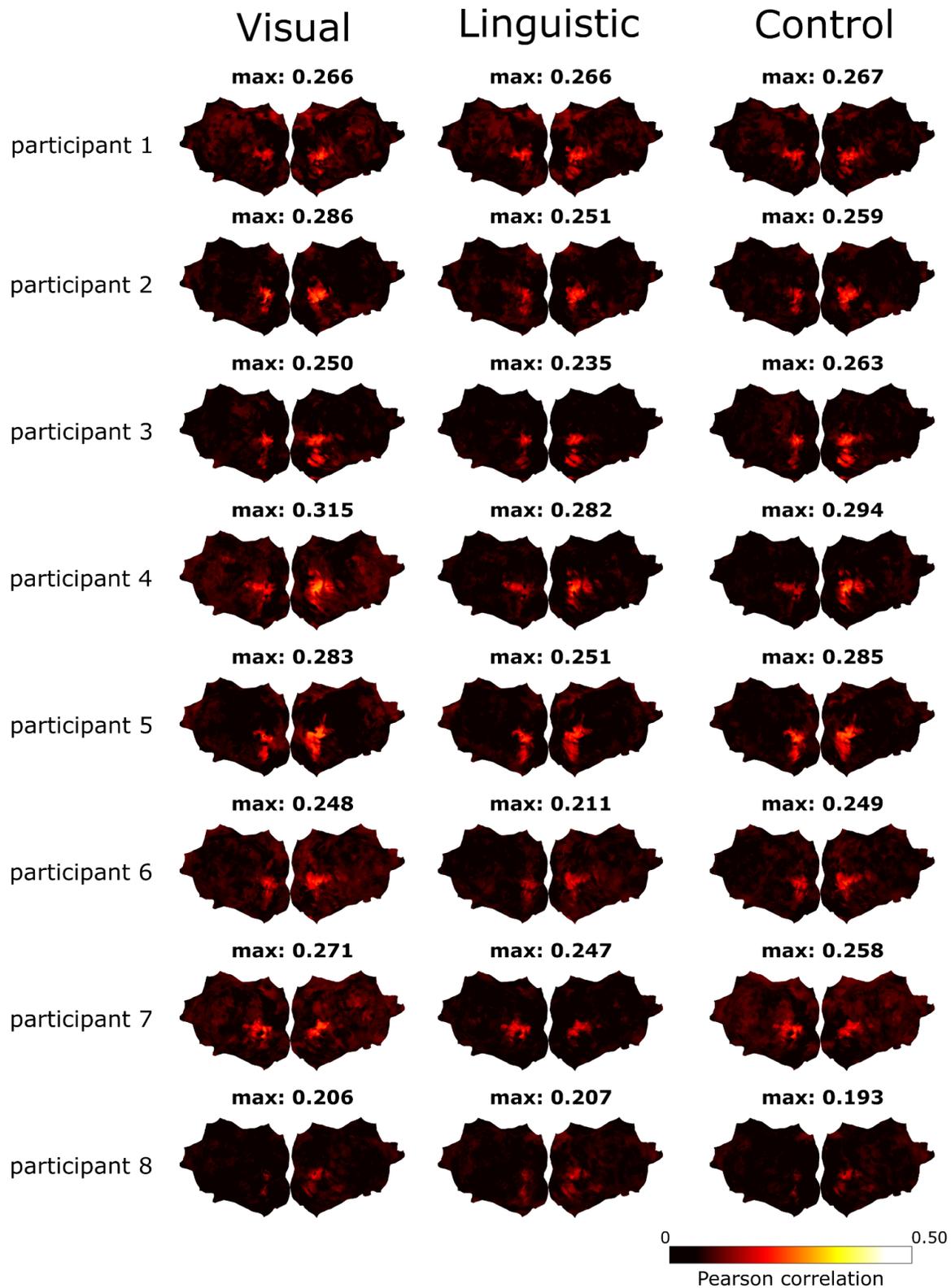

**Supplementary Figure 6. Representational alignment of behavioural similarity judgements in Session 1 and visually evoked brain responses at the participant**

**level.** The averaged modality-specific behavioural RDMs from Session 1 were correlated with brain RDMs using Pearson correlation.

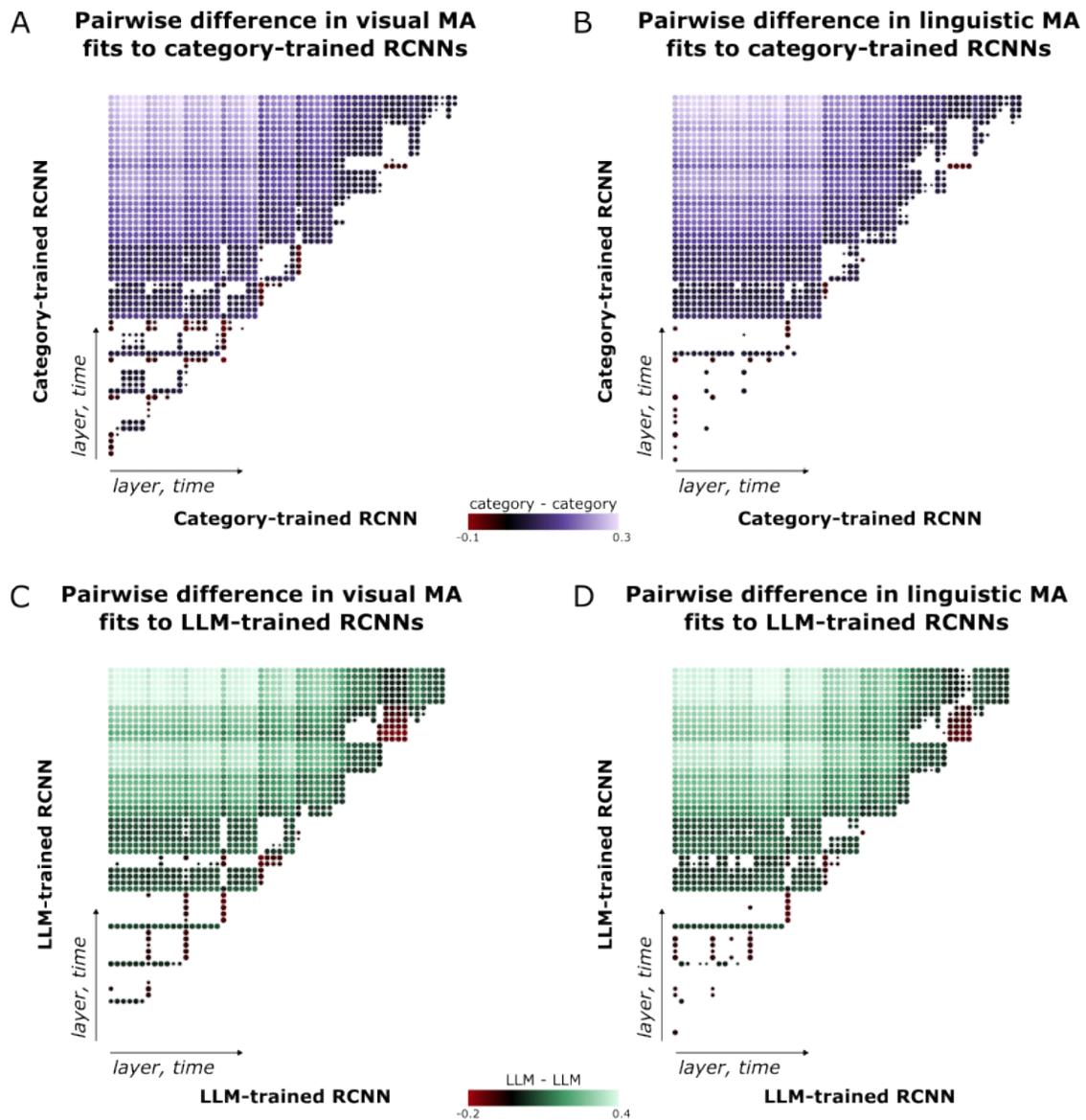

**Supplementary Figure 7. Pairwise difference in behaviour-predicted RCNN RDM fits to observed RCNN RDMs across layers and time steps.** Differences in prediction accuracies were obtained by subtracting Pearson correlations for all pairwise combinations of layers and time steps within an RCNN model type. The colour gradient depicts the size of the difference. The significance of prediction accuracies across seeds within each layer was determined using a one-sided t-test against zero ($p < 0.05$ FDR

corrected). Only significant differences in Pearson correlations are depicted. Radius of the points marks the size of p-values. Comparison between category-trained RCNN RDMs and behaviour-predicted RDMs in vision and language are depicted in panel (A) and (B) respectively. Comparisons between LLM-trained RCNN RDMs and behaviour-predicted RDMs in vision and language are depicted in panel (C) and (D). The layer and time complexity increases from the origin.

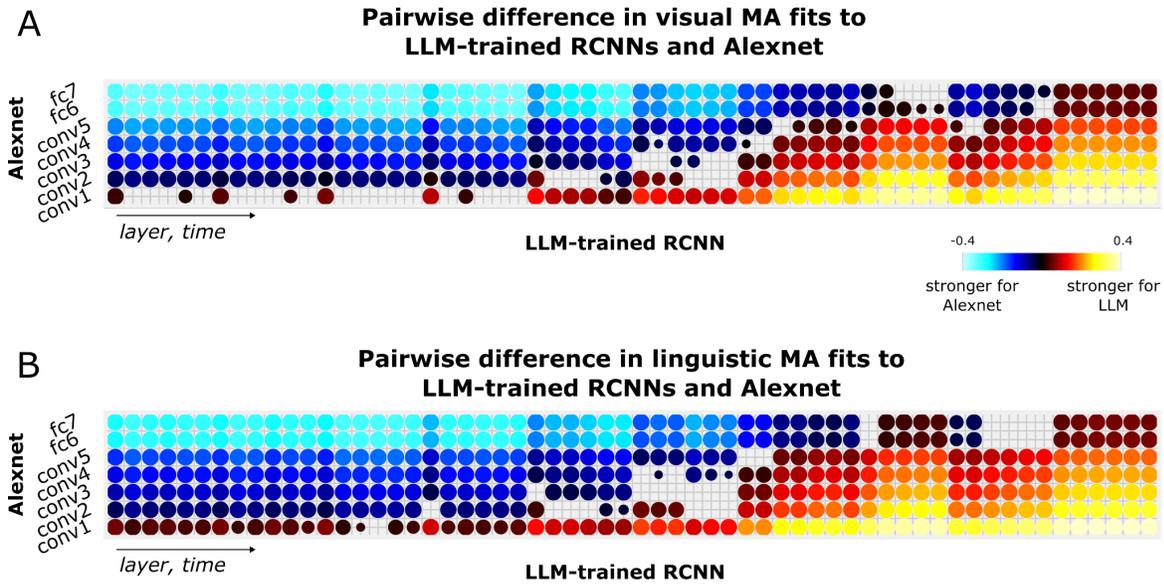

**Supplementary Figure 8. Pairwise difference in behaviour-predicted RCNN RDM fits to observed RCNN and Alexnet RDMs.** The image set was passed through the first five convolutional and two fully-connected layers of Alexnet. The behaviour-predicted RDMs were then compared to observed Alexnet layer RDMs to determine their alignment. The figure shows pairwise differences in Pearson correlations between the observed LLM-trained RCNN and AlexNet RDMs with behaviour-predicted RDMs for the visual (A) and linguistic MA (B). Pearson correlations across RCNN seeds were compared to AlexNet correlations averaged across folds using two-sided, one-sample t-test. Radius of the points marks the size of the p-values ($p < 0.05$ FDR corrected). Only significant differences in cosine similarities are depicted. The colour gradient depicts the size of the difference where mean prediction accuracies from Alexnet were subtracted from mean

prediction accuracies from sentence-computable RCNN. The layer and time complexity of the RCNN increases from the origin of the x-axis.

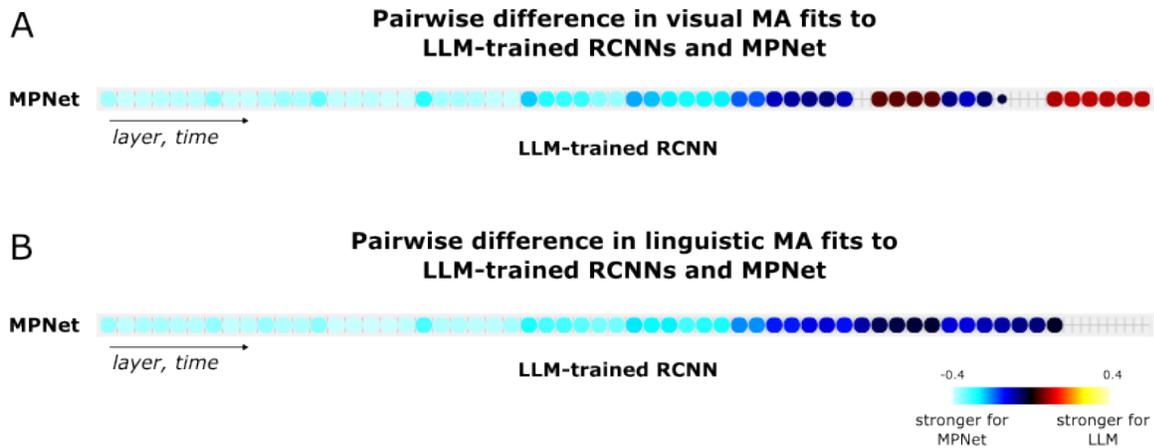

**Supplementary Figure 9. Pairwise difference in behaviour-predicted RCNN RDM fits to observed RCNN and MPNet RDMs.** The figure shows pairwise differences in Pearson correlations between the observed LLM-trained RCNN and MPNet RDMs with behaviour-predicted RDMs for the visual (A) and linguistic (B) modality. The results show a better alignment between the LLM-trained RCNN and the visual modality, but better alignment between the MPNet and the linguistic modality. Pearson correlations across RCNN seeds were compared to MPNet correlations averaged across folds using two-sided, one-sample t-test. Only significant differences in Pearson correlations are depicted. Radius of the points marks the size of the p-values (p < 0.05 FDR corrected). The colour gradient depicts the size of the difference where mean Alexnet trained accuracies were subtracted from mean accuracies of the LMM-trained RCNN. The layer and time complexity increases from the origin of the x-axis.

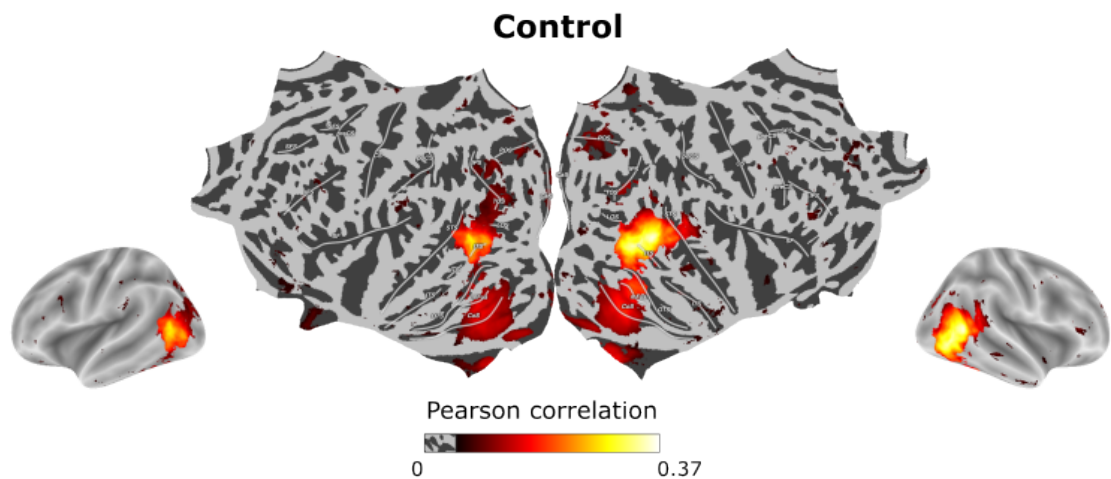

**Supplementary Figure 10. Observed brain RDMs compared to brain RDMs predicted by non-negative least squares of the covert visual condition**. The figure shows Pearson correlations averaged across 10-folds and 8 NSD participants ($p < 0.05$, FDR-corrected, one-sided test). Visual inspection of the surface map reveals a network of activations spanning the occipitotemporal cortex with correlations peaking at 0.327. The active regions are strikingly similar to the activations predicted by the visual and linguistic modality as shown in Figure 2 suggesting a shared representational space regardless of the input modality.

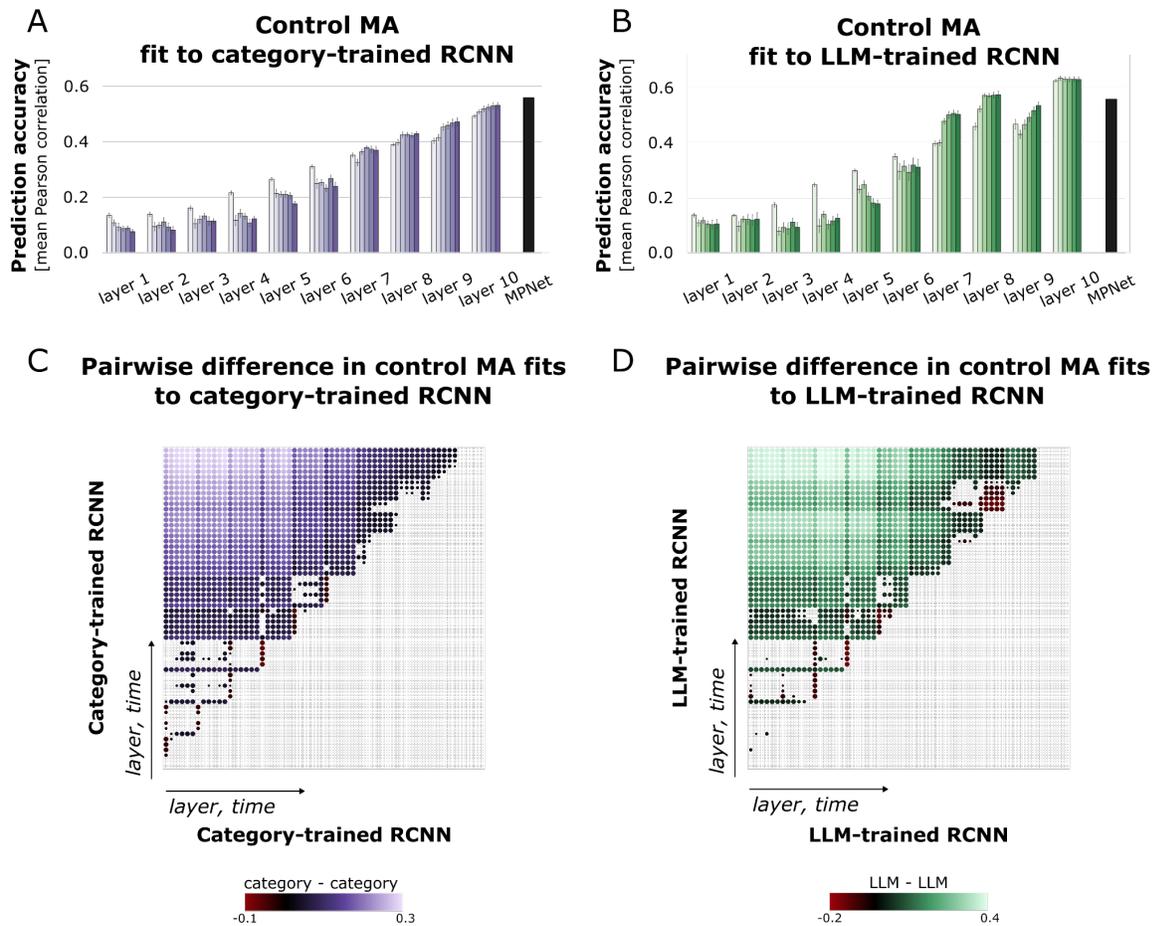

**Supplementary Figure 11. Representational alignment of the behaviour-predicted RDMs from the control MA with category- and LLM-trained RCNNs.** The bars show Pearson correlation between the behaviour-predicted RDM and observed RDM for a given layer and time step, averaged across 10 models trained with different random seeds. The black bar shows averaged Pearson correlations across folds between the behaviour-predicted RDM and observed RDM for the MPNet embeddings of the scene descriptions. The error bars are standard errors of the mean estimated across 10 cross-validation folds for each seed. Prediction accuracy between the category- and LLM-trained RCNN RDMs and behaviour-predicted RDMs for the control MA are depicted in panel (A) and (B) respectively. Differences in prediction accuracies were obtained by subtracting Pearson correlations for all pairwise combinations of layers and time steps within an RCNN model type. The colour gradient depicts the size of the difference. The significance of prediction accuracies across seeds within each layer was determined

using a one-sided t-test against zero ($p < 0.05$ FDR corrected). Only significant differences in Pearson correlations are depicted. Radius of the points marks the size of p-values. Comparison of the category- and LLM-trained RCNN RDMs and behaviour-predicted RDMs for the control MA are depicted in panel (C) and (D respectively. The layer and time complexity increases from the origin.

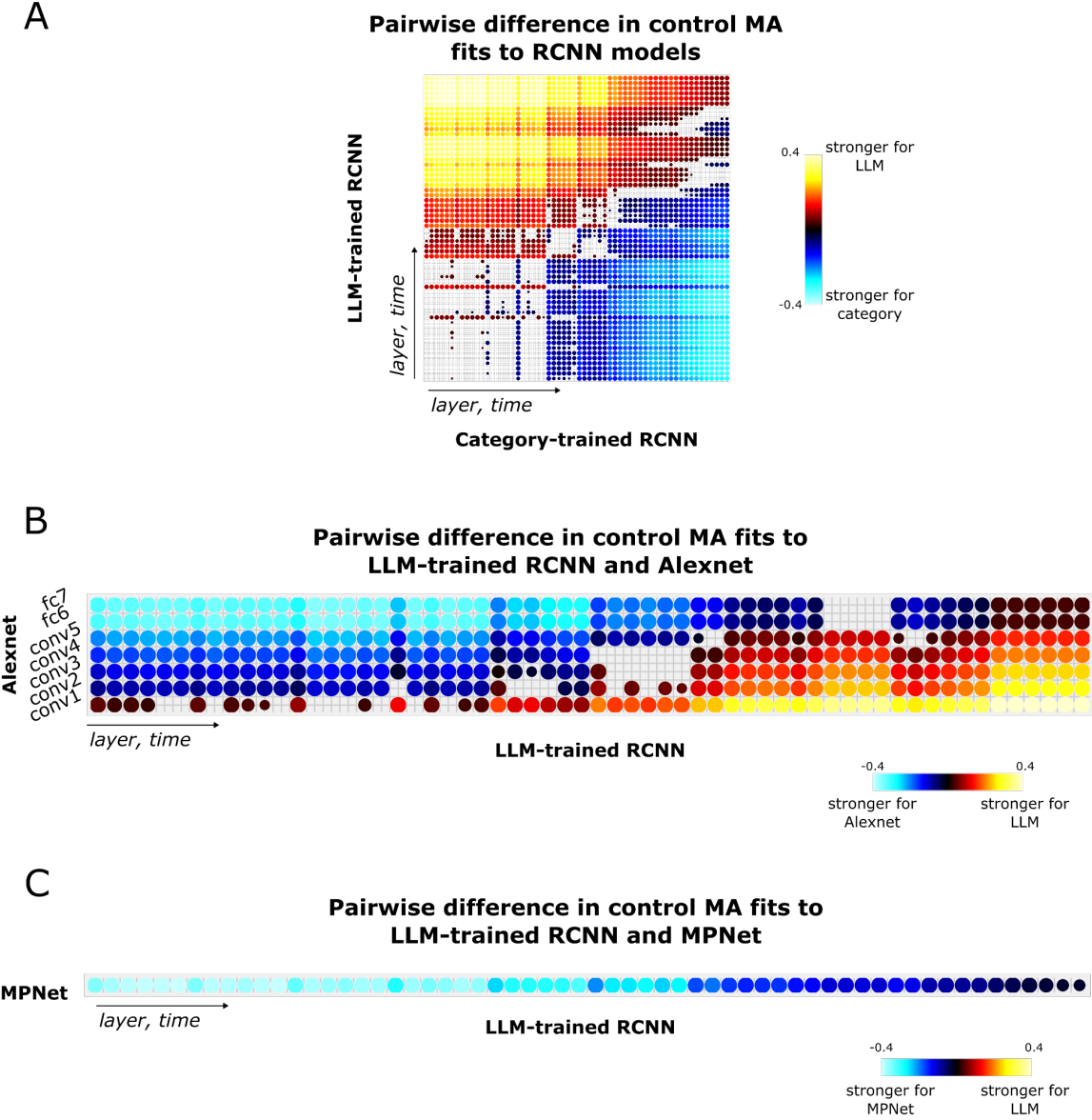

**Supplementary Figure 12. Pairwise difference in behaviour-predicted RCNN RDM fits to observed RCNNs, AlexNet, and MPNet RDMs.** Pairwise difference in Pearson

correlations was calculated to identify a model that best captures the representational structure of the control MA. We compared fits between the behaviour-predicted RDMs for the (A) category- and LLM-trained RCNN, (B) AlexNet, and (C) MPNet. Two-sided t-tests were performed on Pearson correlations across seeds for each layer and time step to determine whether there is a significant difference in model fits with the behaviour-predicted RDMs. Pearson correlations between the LLM-trained RCNN and the behaviour-predicted RCNN RDM across seeds were compared to averaged Pearson correlations from the AlexNet and MPNet using two-sided, one sample t-test. Only significant differences are depicted. Radius of the points marks the size of the p-values ($p < 0.05$ FDR corrected). The colour gradient depicts the size of the difference where mean Pearson correlations of the category-trained RCNN, AlexNet, or MPNet were subtracted from mean prediction accuracies of the LLM-trained RCNN. The layer and time complexity increases from the origin.